\documentclass[12pt]{article}
\usepackage[english,german,french,polish]{babel}
\usepackage[T1]{fontenc}

\begin{document}
\selectlanguage{english}

\baselineskip 0.76cm
\topmargin -0.4in
\oddsidemargin -0.1in

\let\ni=\noindent

\renewcommand{\thefootnote}{\fnsymbol{footnote}}

\newcommand{\SM}{Standard Model }

\pagestyle {plain}

\setcounter{page}{1}

%\pagestyle{empty}

%\addtocounter{equation}{+1}

~~~~~~
\pagestyle{empty}

\begin{flushright}
IFT-- 05/10
\end{flushright}

\vspace{0.4cm}

{\large\centerline{\bf Intrinsically acting Pauli principle as the origin}}

\vspace{0.2cm}

{\large\centerline{\bf of three Standard Model generations of leptons and quarks}}

\vspace{0.5cm}

{\centerline {\sc Wojciech Kr\'{o}likowski}}

\vspace{0.23cm}

{\centerline {\it Institute of Theoretical Physics, Warsaw University}}

{\centerline {\it Ho\.{z}a 69,~~PL--00--681 Warszawa, ~Poland}}

\vspace{0.3cm}

{\centerline{\bf Abstract}}

\vspace{0.2cm}

We turn back to the hypothesis that the Pauli principle, acting intrinsically within leptons 
and quarks, is the origin of their three generations. The adequate formalism is based on 
the generalized Dirac equations arising (in the interaction-free case) from the Klein-Gordon 
equations through the familiar squared-root procedure (but applied in the generic way). 
This leads to the existence of {\it additional} Dirac bispinor indices decoupled from the 
Standard Model gauge fields, thus nonobserved in these fields and, in consequence, {\it not distinguishable} from each other. They are treated as dynamical degrees of freedom 
obeying the Pauli principle along with Fermi statistics. Then, they produce within 
leptons and quarks the total additional spin equal to zero, and cause the existence of 
{\it three and only three} generations of Standard Model leptons and quarks. In the second 
part of the note we discuss the role of the new generation-weighting factors in building up 
the spectra of charged leptons and neutrinos.  

\vspace{0.6cm}

\ni PACS numbers: 12.50.Ch , 12.90+b , 12.10.Dm , 12.10.Kt

\vspace{0.6cm}

\ni April 2005

\vfill\eject

~~~~~
\baselineskip 0.76cm

\pagestyle {plain}

\setcounter{page}{1}

\vspace{0.2cm}

The question, why there are three \SM generations of leptons and quarks, is fundamental in particle physics and so, in contemporary physics as a whole. With some luck, this question, though it is essentially a part of the profound problem of the origin of particle mass, may be simple enough to
get solved independently, perhaps, as a step toward a proper formulation of the mass problem. It is worthwhile to notice that from the methodological point of view the problem of mass introduced already  by Newton was not changed essentially from his time (even by the general theory of relativity): particle mass still has the status of a phenomenological parameter (in Higgs mechanism such a role is played by Yukawa coupling constant).

In this note we turn back to the idea [1] that, in fact, the Pauli principle (along with Fermi statistics) bears responsibility for restricting to three the number of lepton and quark generations. In such a case, leptons and quarks ought to be (in a sense) composite in order to provide some additional degrees of freedom, subject to the restricting action of Fermi statistics. However, according to our idea [1], leptons and quarks {\it are not} composite states of some spatial spin-1/2 preons. Instead, beside the familiar Dirac bispinor index, they {\it get} some {\it additional} Dirac bispinor indices, treated as dynamical degrees of freedom obeying the Pauli principle along with Fermi statistics. Due to this statistics, the additional bispinor indices produce the total additonal spin equal to zero or one half, and can appear only in five configurations 0,2,4 or 1,3, respectively. So, for leptons and quarks (as fermions), they produce the total additional spin 0 and can appear only in three configurations 0,2,4. Hence, {\it three and only three} lepton and quark generations can be realized. 

Thus, in our construction, the familiar notion of spatial compositeness is replaced by the new notion of algebraic compositeness which arises in an act of abstraction from the previous spatial notion. One can see an analogy of this act of algebraic abstraction with the famous Dirac's act of abstraction that has led to the new algebraic notion of spin 1/2 from the familiar spatial notion of orbital angular momentum. If this analogy has a fundamental character, there are no spatial spin-1/2 preons composing leptons and quarks, much like there is no correct spatial model of spin 1/2. If this analogy works practically, rather than on a fundamental level, our Dirac bispinor indices involved in leptons and quarks may unveil only the visible summit of an iceberg of some hidden spatial preonic structure of leptons and quarks.

In order to realize the above idea, we used the dynamical model of generalized Dirac particle [1]. To this end, we postulated the generalized Dirac equations reading (in the interaction-free case) as follows:

%rownanie 1
\begin{equation}
\left( \Gamma^{(N)}\cdot p -  M^{(N)}\right) \psi^{(N)}(x) = 0 \,,
\end{equation}

\ni where the Dirac-type matrices get the form

%rownanie 2
\begin{equation}
\Gamma^{(N)}_\mu \equiv \frac{1}{\sqrt{N}} \sum^N_{i=1}  \gamma^{(N)}_{i \mu} 
\end{equation}

\ni built up linearly from $N$ elements of the Clifford algebra:

%rownanie 3
\begin{equation}
\left\{ \gamma^{(N)}_{i \mu}\,,\,\gamma^{(N)}_{j \nu} \right\} = 2 \delta_{i j} g_{\mu \nu}\,. 
\end{equation}

\ni Here, $N = 1,2,3, ... $, $\;i,j = 1,2, ..., N$ and $\mu, \nu = 0,1,2,3$. From Eqs. (2) and (3) we got for any $N$ the Dirac algebra:

%rownanie 4
\begin{equation}
\left\{ \Gamma^{(N)}_\mu\,,\,\Gamma^{(N)}_\nu \right\} = 2 g_{\mu \nu}\,.
\end{equation}

\ni Thus, $(\Gamma^{(N)}\cdot p)^2 = p^2$, and the Dirac square-root procedure $\sqrt{p^2} 
\rightarrow \Gamma^{(N)}\cdot p$ works in a generic way, leading (in the interaction-free case) from the Klein-Gordon equations $(p^2 - M^{(N)\,2})\psi^{(N)}(x) = 0$ to the generalized Dirac equations (1). Writing $\gamma^{(N)}_{i \mu} = \left(\gamma^{(N)}_{i \mu\,\alpha_1 \alpha_2\ldots \alpha_N\,\beta_1 \beta_2\ldots \beta_N} \right)$, one can see from Eq. (1) that $\psi^{(N)}(x) =  \left(\psi^{(N)}_{\alpha_1\,\alpha_2\,...\alpha_N}(x)\right)$, where each $\alpha_i = 1,2,3,4\;(i = 1,2,...,N)$ is a Dirac bispinor index in the chiral representation.

For $ N = 1$ Eq. (1) is evidently the usual Dirac equation, for $ N = 2$ it is known as the Dirac form [2] of K\"{a}hler equation [3], while for $N \geq 3$ Eq. (1) gives us new generalized Dirac equations [1]. They describe some spin-halfinteger and spin-integer particles for $N$ odd and $N$ even, respectively.

The Dirac-type matrices $\Gamma^{(N)}_\mu $ for any $N$ can be embedded into the new Clifford algebra:

%rownanie 5
\begin{equation}
\left\{ \Gamma^{(N)}_{i \mu}\, ,  \,\Gamma^{(N)}_{j \nu} \right\} = 2\delta_{i j} g_{\mu \nu}\;,
\end{equation}

\ni isomorphic to the previous Clifford algebra (3) of $\gamma^{(N)}_{i \mu}$, where the new elements $\Gamma^{(N)}_{i \mu}$ are defined by the properly normalized Jacobi linear combinations of $\gamma^{(N)}_{i \mu}$:

%rownanie 6
\begin{eqnarray}
\Gamma^{(N)}_{1 \mu} & \equiv & \Gamma^{(N)}_\mu \equiv \frac{1}{\sqrt{N}} \left(\gamma^{(N)}_{1 \mu} + \ldots + \gamma^{(N)}_{N \mu} \right) \,, \nonumber \\ \Gamma^{(N)}_{i \mu} & \equiv & \frac{1} {\sqrt{i(i - 1)}} \left[ \gamma^{(N)}_{1 \mu} + \ldots + \gamma^{(N)}_{i\!-\!1\, \mu} - (i - 1) \gamma^{(N)}_{i \mu} \right]\;\; (i = 2,...,N)\;.
\end{eqnarray}

\ni Thus $\Gamma^{(N)}_{1 \mu}$ and $\Gamma^{(N)}_{2 \mu}, ..., \Gamma^{(N)}_{N \mu}$, respectively, present the "centre-of-mass"\,and "relative"\,Dirac-type matrices. Note that for any $N$ the generalized Dirac equation (1) does not involve the "relative"\,Dirac-type matrices $\Gamma^{(N)}_{2 \mu}, ..., \Gamma^{(N)}_{N \mu}$, including solely the "centre-of-mass"\,Dirac-type matrices $\Gamma^{(N)}_{1 \mu} \equiv \Gamma^{(N)}_{ \mu}$.

It is not difficult to see that for any $N$ the total spin tensor is given as

%rownanie 7
\begin{equation}
\sum^N_{i=1}  \sigma^{(N)}_{i \mu \nu} = \sum^N_{i=1}  \Sigma^{(N)}_{i \mu \nu}  \,,
\end{equation}

\ni where

%rownanie 8
\begin{equation} 
\sigma^{(N)}_{j \mu \nu} \equiv \frac{i}{2} \left[ \gamma^{(N)}_{j \mu} \, , \, \gamma^{(N)}_{j \nu} \right] \;\; ,\;\; \Sigma^{(N)}_{j \mu \nu} \equiv \frac{i}{2} \left[ \Gamma^{(N)}_{j \mu} \, , \, \Gamma^{(N)}_{j \nu} \right] \,. 
\end{equation}

\ni The total spin tensor (7) becomes the generator of Lorentz transformations for $\psi^{(N)}(x)$.

Now, it is convenient to use for any $N$ the chiral representation of Jacobi-type Clifford matrices $\Gamma^{(N)}_{i \mu} = \left( \Gamma^{(N)}_{i \mu\,\alpha_1 \alpha_2\ldots \alpha_N\,\beta_1 \beta_2\ldots \beta_N}\right)$ in place of the chiral representation of individual Clifford matrices $\gamma^{(N)}_{i \mu} = \left( \gamma^{(N)}_{i \mu\,\alpha_1 \alpha_2\ldots \alpha_N\,\beta_1 \beta_2\ldots \beta_N}\right)$. Then, one may choose

%rownanie 9
\begin{equation}
\Gamma^{(N)}_{1 \mu} \equiv \Gamma^{(N)}_\mu =  \gamma_\mu \otimes \underbrace{{\bf 1}\otimes \cdots \otimes {\bf 1}}_{N-1 \;{\rm times}}  \;,
\end{equation}

\ni where $\gamma_\mu$ and {\bf 1} are the usual $4\times 4$ Dirac matrices. In this new chiral representation the generalized Dirac equations (1) for $\psi^{(N)}_{i \mu}(x) = \left( \psi^{(N)}_{i \mu \alpha_1 \alpha_2 \ldots \alpha_N}(x)\right)$ take the forms

%rownanie 10
\begin{equation}
\left( \gamma \cdot p  - M^{(N)}\right)_{\alpha_1\beta_1} \psi^{(N)}_{\beta_1 \alpha_2 \ldots \alpha_N}(x) = 0\;,
\end{equation}

\ni where $\alpha_1$ and $\alpha_2 , \ldots, \alpha_N$ are the "centre-of-mass"\,and \,"relative"\, Dirac bispinor indices, respectively (the latter appear for $N>1$). Note that in the generalized Dirac equations (10)  the  "relative"\, Dirac bispinor indices are free from any coupling, but still are subject to Lorentz transformations. 

The Standard Model gauge interactions can be introduced to the generalized Dirac equations (10)  by means of the minimal substitution $p \rightarrow p - g A(x)$, where $p$ plays the role of the 
"centre-of-mass"\, four-momentum and so, $x$ --- the role of "centre-of-mass"\, four-position. Then,

%rownanie 11
\begin{equation}
\left[ \gamma \cdot \left( p - g A(x)\right) - M^{(N)}\right] \psi^{(N)}(x) = 0\;,
\end{equation}

\ni where $g\gamma \cdot A(x)$ symbolizes the \SM  gauge coupling involving within $A(x)$ the familiar weak-isospin   and color matrices, the weak-hypercharge dependence as well as the usual $4\times 4$ Dirac chiral matrix $\gamma_5 \equiv i \gamma_0 \gamma_1 \gamma_2 \gamma_3$. 

In Eq. (11) the Standard Model gauge fields interact only with the "centre-of-mass" \,index $\alpha_1$ that, therefore, is distinguished from the  "relative" \,indices, nonobserved in these fields and, in consequence, not distinguishable from each other. This was the reason, why some years ago we conjectured that the "relative" \,Dirac bispinor  indices $\alpha_2, \ldots, \alpha_N$ are all indistinguishable dynamical objects obeying the Pauli principle along with Fermi statistics requiring the full antisymmetry of wave function $\psi^{(N)}_{\alpha_1 \alpha_2 \ldots \alpha_N}(x)$ with respect to the indices $\alpha_2, \ldots, \alpha_N$ [4, 1]. Hence, due to this Pauli principle (realized intrinsically), only five values of $N$ satisfying the condition $N-1\leq 4$ are allowed, namely $N = 1,3,5$ for $N$ odd and $N = 2,4$ for $N$ even. Then, from the postulate of relativity and the probabilistic interpretation of $\psi^{(N)}(x) \equiv \left(\psi^{(N)}_{\alpha_1 \alpha_2 \ldots \alpha_N}(x)\right) $ we were able to infer that $N$ odd and $N$ even correspond to states with total spin 1/2 and total spin 0, respectively [4, 1].

Thus, the generalized Dirac equation (11), jointly with the Pauli principle (realized intrinsically),  justifies the existence in Nature of {\it three and only three} generations of  leptons and quarks. In addition, there should exist {\it two and only two} generations of spin-0 fundamental bosons 
(weak-isospin doublets and singlets, and colored singlets and triplets) also coupled to the \SM gauge bosons. Note that the lack of one generation of these spin-0 bosons makes the construction of  three-generation supersymmetric theory impossible..

The wave functions or fields of spin-1/2 fundamental fermions (leptons and quarks) of three generations $N = 1,3,5$ can be written down in terms of $\psi^{(N)}_{\alpha_1 \alpha_2 \ldots \alpha_N}(x)$ as follows [4, 1]:

%rownanie 12
\begin{eqnarray} 
\psi^{(f_1)}_{\alpha_1}(x) & = & \psi^{(1)}_{\alpha_1}(x) \;, \nonumber \\
\psi^{(f_2)}_{\alpha_1}(x) & = & \frac{1}{4}\left(C^{-1} \gamma_5 \right)_ {\alpha_2 \alpha_3} \psi^{(3)}_{\alpha_1 \alpha_2 \alpha_3}(x) = \psi^{(3)}_{\alpha_1 1 2}(x) = \psi^{(3)}_{\alpha_1 3 4}(x) \;,\nonumber \\
\psi^{(f_3)}_{\alpha_1}(x) & = & \frac{1}{24}\varepsilon_{\alpha_2 \alpha_3 \alpha_4 \alpha_5} \psi^{(5)}_{\alpha_1 \alpha_2 \alpha_3 \alpha_4 \alpha_5}(x) = \psi^{(5)}_{\alpha_1 1 2 3 4}(x) \;,
\end{eqnarray}  

\ni where $ \psi^{(N)}_{\alpha_1 \alpha_2 \ldots \alpha_N}(x)\;(n=1,3,5)$ carry also the \SM (composite) label (suppressed in our notation), while $C$ denotes the usual $4\times 4$ charge-conjugation matrix.  Writing explicitly, one gets $f_1 = \nu_e, e^-, u, d \;,\; f_2 = \nu_\mu, \mu^-, c, s $ and $f_3 = \nu_\tau, \tau^-, t , b $, thus each $f_i\;(i=1,2,3)$ carries the same suppressed \SM (composite) label. One can see that, due to the full antisymmetry in $\alpha_2, ..., \alpha_N $ indices, the wave functions or fields (12) corresponding to $N =1,3,5$ appear (up to the sign) with the multiplicities 1, 4, 24,  respectively. Thus, there is defined the generation-weighting matrix [4, 1]

%rownanie 13
\begin{equation} 
\rho^{1/2} = {\rm diag}(\rho^{1/2}_1,\rho^{1/2}_2,\rho^{1/2}_3) =  \frac{1}{\sqrt{29}}  \left( \begin{array}{ccc}  1  & 0 & 0 \\ 0 & \sqrt4 & 0  \\ 0 & 0 & \sqrt{24} \end{array} \right) \;, 
\end{equation}

\ni where Tr$\rho = 1$. It gives

%rownanie 14
\begin{equation}
\psi^{(N_i)\dagger}(x) \psi^{(N_i)}(x) = 29 \rho_i \psi^{(f_i)\dagger}(x) \psi^{(f_i)}(x)  
\end{equation}

\ni for $N_i =1,3,5$ and $i = 1,2,3$, where $\rho_i =1/29, 4/29, 24/29$.

One may ask an interesting and in practice important question as to whether experimental lepton and quark mass spectra can be built up efficiently from the numbers $N_i = 1,3,5$ numerating generations and from the generation-weighting factors $\rho_i =1/29, 4/29, 24/29$, $(i = 1,2,3)$. Some years ago, we obtained a positive answer to it in the case of charged leptons $e_i = e^-, \mu^-, \tau^- $ [4, 1]. In fact, the mass formula

%rownanie 15 (Leger}
\begin{equation}
m_{e_i} = \mu^{(e)} \rho_i  \left(N^2_i + \frac{\varepsilon^{(e)} - 1}{N^2_i} \right) 
\end{equation}

\ni with $\mu^{(e)} > 0$ and $\varepsilon^{(e)} > 0$ denoting free constants, when rewritten explicitly as

%rownanie 16
\begin{equation}
m_e = \frac{\mu^{(e)}}{29} \varepsilon^{(e)} \;,\;  m_\mu = \frac{\mu^{(e)}}{29} \frac{4}{9} (80 +  \varepsilon^{(e)}) \;,\;  m_\tau = \frac{\mu^{(e)}}{29} \frac{24}{25} (624 + \varepsilon^{(e)})\;,
\end{equation}

\ni leads to the prediction

%rownanie 17
\begin{equation}
m_\tau = \frac{6}{125}(351 m_\mu - 136 m_e) = 1776.80 \;{\rm MeV}
\end{equation}

\ni and determines both constants

%rownanie 18
\begin{equation}
\mu^{(e)} = \frac{29(9m_\mu - 4m_e)}{320} = 85.9924 \;{\rm MeV} \;,\; \varepsilon^{(e)} = \frac{320 m_e}{9m_\mu - 4m_e} = 0.172329 \,.
\end{equation}

\ni Here, the experimental values $ m_e = 0.510999$ MeV and  $ m_\mu = 105.658$ MeV are used as an input. The prediction (17) is really close to the experimetal value $ m_\tau = 1776.99^{+0.29}_{-0.26}$ MeV [5]. 

Recently, we considered an analogical question for neutrino mass states $\nu_i = \nu_1, \nu_2, \nu_3 $ related to active neutrinos $ \nu_e, \nu_\mu, \nu_\tau $ [6]. Since mass neutrinos display experimentally a less hierarchical spectrum than charged leptons, namely

%rownanie 19
\begin{equation}
7.2 < \Delta m^2_{21}/(10^{-5}\;{\rm eV}^2) < 9.1\;,\; 1.9 < \Delta m^2_{32}/(10^{-3}\;{\rm eV}^2) < 3.0 
\end{equation}

\ni with the best fits

%rownanie 20
\begin{equation} 
\Delta m^2_{21} \sim 8.1\times 10^{-5}\;{\rm eV}^2 \;,\; \Delta m^2_{32} \sim 2.4 \times10^{-3} \;{\rm eV}^2 
\end{equation}

\ni where $\Delta m^2_{ji} \equiv m^2_{\nu_j} - m^2_{\nu_i}$ [7], we used only the generation-weighting factors $\rho_i $, ignoring the numbers $N_i$ numerating generations. 

In the simplest case such a mass formula is

%rownanie 21
\begin{equation}
m_{\nu_i} = \mu^{(\nu)} \rho_i 
\end{equation}

\ni with $\mu^{(\nu)} > 0$ being a free constant. This implies that

%rownanie 22
\begin{equation}
m_{\nu_1} : m_{\nu_2} : m_{\nu_3}  = 1 : 4 : 24
\end{equation}

\ni and

%rownanie 23
\begin{equation}
\mu^{(\nu)} = m_{\nu_1} + m_{\nu_2} + m_{\nu_3} = 29m_{\nu_1} = \frac{29}{4} m_{\nu_2}  = \frac{29}{24} m_{\nu_3} \,. 
\end{equation}

\ni From Eq. (22)

%rownanie 24
\begin{equation}
\Delta m^2_{32}/\Delta m^2_{21} = \frac{112}{3} \simeq 37 \,,
\end{equation}

\ni while the experimental estimates (20) give $\Delta m^2_{32}/\Delta m^2_{21} \sim 30$. Using the experimental estimates (19 ) and (20) for $\Delta m^2_{21}$, one gets from Eq. (24) the predictions

%rownanie 25
\begin{equation}
2.7 < \Delta m^2_{32}/(10^{-3}\;{\rm eV}^2) < 3.4 
\end{equation}

\ni and 

%rownanie 26
\begin{equation}
\Delta m^2_{32} \sim 3.0 \times 10^{-3}\;{\rm eV}^2 
\end{equation}

\ni as well as

%rownanie 27
\begin{equation}
m_{\nu_1} \sim 2.3\times 10^{-3}\;{\rm eV}\,,\, m_{\nu_2} \sim 9.3\times 10^{-3}\;{\rm eV}\,,\, m_{\nu_3} \sim 5.6\times 10^{-2}\;{\rm eV}
\end{equation}

\ni and 

%rownanie 28
\begin{equation}
\mu^{(\nu)} \sim 6.7\times 10^{-2}\;{\rm eV} \,.
\end{equation}

\ni The values (25) and (26) are too large, though of the correct order.

Good predictions are given by the two-parameter mass formula

%rownanie 29
\begin{equation}
m_{\nu_i} = \mu^{(\nu)} \rho_i (1 - \beta \delta _{i3}) 
\end{equation}

\ni with $\beta > 0$ denoting a second free parameter. This leads to

%rownanie 30
\begin{equation}
m_{\nu_1} : m_{\nu_2} : m_{\nu_3} = 1 : 4 : 24(1 - \beta)
\end{equation}

\vspace{-0.3cm}

\ni and

\vspace{-0.3cm}

%rownanie 31
\begin{equation}
\mu^{(\nu)} = \frac{29}{5+ 24(1-\beta)}(m_{\nu_1} + m_{\nu_2} + m_{\nu_3}) = 29m_{\nu_1} = \frac{29}{4} m_{\nu_2}  = \frac{29}{24(1-\beta)} m_{\nu_3} \,. 
\end{equation}

\vspace{0.2cm}

\ni From Eq. (30)

\vspace{-0.2cm}

%rownanie 32
\begin{equation}
\Delta m^2_{32}/\Delta m^2_{21} =  \frac{16[36(1-\beta)^2 - 1]}{15} \;,
\end{equation}

\vspace{0.2cm}

\ni what gives the value $\Delta m^2_{32}/\Delta m^2_{21} \sim 30$ consistent with the experimental estimates (20) if

\vspace{-0.2cm}

%rownanie 33
\begin{equation}
\beta \sim 0.10 \;.
\end{equation}

\ni So, $\beta $ is a small parameter. Using the experimental estimates (19 ) and (20) for $\Delta m^2_{21}$, one obtains from Eqs. (32) and (33) that

%rownanie 34
\begin{equation}
2.2 < \Delta m^2_{32}/(10^{-3}\;\, {\rm eV}^2) < 2.7 
\end{equation}

\vspace{-0.2cm}

\ni and

\vspace{-0.2cm}

%rownanie 35
\begin{equation}
\Delta m^2_{32} \sim 2.4 \times 10^{-3}\;\, {\rm eV}^2 
\end{equation}

\ni as well as

%rownanie 36
\begin{equation}
m_{\nu_1} \sim 2.3 \times 10^{-3}\;\, {\rm eV} \,,\, m_{\nu_2} \sim 9.2 \times 10^{-3}\;\, {\rm eV} \,,\, m_{\nu_3} \sim 5.0 \times 10^{-2}\;\, {\rm eV} 
\end{equation}

\ni and 

%rownanie 37
\begin{equation}
\mu^{(\nu)} \sim 6.7 \times 10^{-2}\;\, {\rm eV} \;.
\end{equation}

\ni Here, the experimental estimates (20) for $\Delta m^2_{21}$ and $\Delta m^2_{32}$ are both the input. But, one of three neutrino masses $m_{\nu_i}$ is still a prediction.

The mass formulae for up and down quarks built up along analogical lines were considered in Ref. [8] (there also, an additional parameter was introduced for $i=3$, as in Eq. (29)).

Alternatively, we may consider the possibility that the neutrino mass formula depends not only on the lepton generation-weighting factors $\rho_i $ but also, very weakly, on the number $N_i $ numerating generations. Then, for example,

\vspace{-0.2cm}

%rownanie 38
\begin{equation}
m_{\nu_i} = \mu^{(\nu)} \rho_i (1 - \eta N^2_i) 
\end{equation}

\ni where a second free parameter $\eta > 0$ is expected to be small. This gives

%rownanie 39
\begin{equation}
m_{\nu_1} : m_{\nu_2} : m_{\nu_3} = 1-\eta : 4(1 - 9\eta : 24(1 - 25\eta)
\end{equation}

\vspace{-0.3cm}

\ni and

\vspace{-0.3cm}

%rownanie 40
\begin{equation}
\mu^{(\nu)} = \frac{29}{29 - 637\eta}(m_{\nu_1} + m_{\nu_2} + m_{\nu_3}) = \frac{29}{1-\eta}m_{\nu_1} = \frac{29}{4(1-9\eta)} m_{\nu_2}  = \frac{29}{24(1-25\eta)} m_{\nu_3} \,. 
\end{equation}

\vspace{0.2cm}

\ni From Eq. (39)

\vspace{-0.2cm}

%rownanie 41
\begin{equation}
\Delta m^2_{32}/\Delta m^2_{21} =  \frac{16[36(1-25\eta)^2 - (1-9\eta)^2]}{16(1-9\eta)^2 - (1 - \eta)^2} \;,
\end{equation}

\vspace{0.2cm}

\ni implying the experimentally consistent value $\Delta m^2_{32}/\Delta m^2_{21} \sim 30$ if

%rownanie 42
\begin{equation}
\eta \sim 6.1\times 10^{-3} 
\end{equation}

\ni (or $\eta \sim 5.6\times 10^{-2}$). Thus, $\eta $ is really a small parameter ($\Delta m^2_{32}/\Delta m^2_{21} = 112/3 \sim 37$ for $\eta = 0$). With the experimental estimates (19 ) and (20) for $\Delta m^2_{21}$ one gets from Eqs. (41) and (42) that

\vspace{-0.2cm}

%rownanie 43
\begin{equation}
2.2 < \Delta m^2_{32}/(10^{-3}\;\, {\rm eV}^2) < 2.7 
\end{equation}

\vspace{-0.2cm}

\ni and

\vspace{-0.2cm}

%rownanie 44
\begin{equation}
\Delta m^2_{32} \sim 2.4 \times 10^{-3}\;\, {\rm eV}^2 
\end{equation}

\ni as well as

%rownanie 45
\begin{equation}
m_{\nu_1} \sim 2.5 \times 10^{-3}\;\, {\rm eV} \,,\, m_{\nu_2} \sim 9.3 \times 10^{-3}\;\, {\rm eV} \,,\, m_{\nu_3} \sim 5.0 \times 10^{-2}\;\, {\rm eV} 
\end{equation}

\vspace{-0.2cm}

\ni and 

\vspace{-0.2cm}

%rownanie 46
\begin{equation}
\mu^{(\nu)} \sim 7.2 \times 10^{-2}\;\, {\rm eV} \;.
\end{equation}

\ni Here, both experimental estimates (20) for $\Delta m^2_{21}$ and $\Delta m^2_{32}$ are the input. A prediction is one of three neutrino masses $m_{\nu_i}$.

\vfill\eject

\vspace{0.6cm}

{\centerline{\bf References}}

\vspace{0.25cm}

{\everypar={\hangindent=0.65truecm}
\parindent=0pt\frenchspacing

[1]~For a recent preseentation {\it cf. } W. Kr\'{o}likowski, {\it Acta Phys. Pol.} {\bf B 33}, 2559 (2002) ({\tt hep--ph/0201186}). 

\vspace{0.2cm}

[2]~T. Banks, Y. Dothan and D.~~Horn, {\it Phys. Lett.} {\bf B 117}, 413 (1982).

\vspace{0.2cm}

[3]~E. K\"{a}hler, {\it Rendiconti di Matematica} {\bf 21}, 425 (1962); {\it cf.} also D.~Ivanenko and L.~Landau, {\it Z. Phys.} {\bf 48}, 341 (1928).

\vspace{0.2cm}

[4]~W. Kr\'{o}likowski, {\it Acta Phys. Pol.} {\bf B 21}, 871 (1990); {\it Phys. Rev.} {\bf D 45}, 3222 (1992); {\it Phys. Rev.} {\bf D 46}, 5188 (1992); in {\it Spinors, Twistors, Clifford Algebras and Quantum Deformations (Proc. 2nd Max Born Symposium 1992)}, eds. Z.~Oziewicz {\it et al.}, Kluwer 1993; {\it Acta Phys. Pol.} {\bf B 24}, 1149 (1993); {\it Acta Phys. Pol.} {\bf B 27}, 2121 (1996); {\it cf.} also Appendices in {\it Acta Phys. Pol.} {\bf B 32}, 2961 (2001) ((\tt hep-ph/0108157}) and {\it Acta Phys. Pol.} {\bf B 33}, 1747 (2002) ({\tt hep-ph/0201004v2}). 

\vspace{0.2cm}

{\everypar={\hangindent=0.65truecm}
\parindent=0pt\frenchspacing

[5]~Particle Data Group, {\it Review of Particle Physics, Phys.~Lett.} {\bf B 592} (2004).

\vspace{0.2cm}
 
[6]~W. Kr\'{o}likowski, {\tt hep--ph/0503074}). 

\vspace{0.2cm}

[7]~For a review {\it cf. } J.W.F.~Valle, {\tt hep--ph/0410103}; D. Kie{\l}czewska, {\it Nucl. Phys.} {\bf B} ({\it Proc. Suppl.}) {\bf 136}, 77 (2004).

\vspace{0.2cm}

[8]~W. Kr\'{o}likowski, {\it Acta Phys. Pol.} {\bf B 35}, 673 (2004); {\it cf.} also {\it Acta Phys. Pol.} {\bf B 35}, 2095 (2004) ({\tt hep--ph/0401101}).

\vfill\eject

\end{document}